\documentclass[11pt,a4paper]{article}
\pdfoutput=1
\usepackage{jcappub}

\usepackage{hyperref}
\usepackage{tabularx,afterpage,xifthen}
\usepackage{amsmath,amsfonts,amssymb,mathtools}
\usepackage{slashed}
\usepackage{comment}
\usepackage{tikz}
\usepackage{caption}
\usepackage{subcaption}
\usepackage{graphicx, xcolor}

\newcommand\rmd{{\rm d}}
\newcommand\fd{f_{\rm D}}

\newcommand\mpd{m_{\rm p_{D}}}
\newcommand\med{m_{\rm e_{D}}}
\newcommand\alphad{\alpha_{\rm D}}
\newcommand\xid{\xi_{\rm D}}

\newcommand\kpeak{k_{\rm{peak}}}
\newcommand\hpeak{h_{\rm{peak}}}
\newcommand\kdamp{k_{\rm{damp}}}
\newcommand\kdao{k_{\rm{DAO}}}

\newcommand{\classadm}{\texttt{CLASS-aDM}}
\newcommand\lcdm{\Lambda \text{CDM}}

\title{Constraining Dark Acoustic Oscillations with the high-redshift UV Luminosity Function}

\author{Jared Barron,$^{a}$}
\author{David Curtin,$^{b, c}$}
\author{Hongwan Liu,$^{d}$}
\author{Julian B. Mu\~noz,$^{e,f,g}$}
\author{Sandip Roy$^{h}$}

\affiliation[a]{C.N. Yang Institute for Theoretical Physics,
Stony Brook University, Stony Brook, NY 11794, USA}
\affiliation[b]{Department of Physics, University of Toronto, Toronto, ON M5S 1A7, Canada}
\affiliation[c]{Theoretical Physics Department, CERN, 1211 Geneva, Switzerland}
\affiliation[d]{Physics Department, Boston University, Boston, MA 02215, USA}
\affiliation[e]{Department of Astronomy, The University of Texas at Austin, 2515 Speedway Boulevard, Austin, TX 78712, USA}
\affiliation[f]{Cosmic Frontier Center, The University of Texas at Austin, Austin, TX 78712, USA}
\affiliation[g]{Texas Center for Cosmology \& Astroparticle Physics, Austin, TX 78712, USA}
\affiliation[h]{Department of Astronomy \& Astrophysics, University of California, San Diego, La Jolla, CA 92093, USA}
\emailAdd{jared.barron@stonybrook.edu}

\abstract{Dark acoustic oscillations (DAOs) in the matter power spectrum can arise in many different dark sector models, and can imprint on a variety of cosmological observables. In this work we use measurements of the galactic UV luminosity function (UVLF) at high redshifts to constrain the dark acoustic oscillation feature at small scales in a model-agnostic way. We introduce a phenomenological transfer function model for a dark sector with a species undergoing DAOs which can accommodate sub-dominant dark matter abundances, and obtain constraints on its parameters. In order to predict the UVLF, we employ an Extended Press-Schechter formalism which we calibrate using N-body simulations with initial conditions featuring DAOs. Using measurements from the Hubble Space Telescope, James Webb Space Telescope, Subaru Telescope, and Canada-France-Hawaii Telescope, we constrain the wave number of the first DAO peak to be at $k \gtrsim 50\ h/\mathrm{Mpc}$, unless the fraction of dark matter undergoing DAOs is less than $0.07$.  
}

\begin{document}

\begin{flushright}
    CERN-TH-2025-200
\end{flushright}

\maketitle
    \section{Introduction}

    Evidence from astrophysical and cosmological observations suggests that most of the matter in the Universe is in the form of dark matter, which is mostly invisible, cold, and collisionless~\cite{Sofue:2000jx,Planck:2018vyg}. However, the precise nature of the dark matter remains elusive, and there are many models of more complex dark sectors that satisfy the same basic constraints as CDM, but differ in more subtle ways, particularly on small scales where observational constraints are weaker. A common feature of many dark sector models beyond $\lcdm$ is modification of the growth of density perturbations in the early universe, leading to deviations from the expected scaling behavior of the matter power spectrum $P(k)\propto k^{-3}$ at scales $k > k_{\mathrm{eq}}$. 

    One such class of modifications, which can naturally arise in many well-motivated dark-sector models, is known as dark acoustic oscillations (DAOs), characterized by suppression and damped oscillations in the matter power spectrum at small length scales. The presence of interactions in the dark sector between some or all of the dark matter (DM) and a dark radiation (DR) species in the early Universe, such as in the atomic dark matter (aDM) model, is a natural way to generate DAOs~\cite{Feng:2009mn,Kaplan:2009de,Cyr-Racine:2012tfp,Cyr-Racine:2013fsa}. The pressure support of the dark radiation at early times drives acoustic oscillations in the dark matter to which it couples, in the same way that the pressure support of Standard Model photons leads to baryon acoustic oscillations. Other dark sector paradigms can also lead to DAOs, such as the interacting dark matter (IDM) model with dark matter-baryon interactions \cite{Chen:2002yh,Sigurdson:2004zp,Boehm:2004th,McDermott:2010pa,Dvorkin:2013cea,Gluscevic:2017ywp,Boddy:2018kfv,Xu:2018efh,Boddy:2018wzy,Nadler:2019zrb,DES:2020fxi,Maamari:2020aqz,He:2023dbn,Nadler:2025fcv,Fischer:2025snw}. 
     
    Models of this type have been the subject of great interest, since they have the potential to solve various apparent discrepancies between the predictions of $\lcdm$ and observations that have appeared over the years \cite{Kaplan:2009de,Cyr-Racine:2012tfp,Cyr-Racine:2013fsa,Foot:2013lxa,Foot:2013vna,Foot:2014uba,Randall:2014kta,Vogelsberger:2015gpr,Foot:2015mqa,Buen-Abad:2015ova,Lesgourgues:2015wza,Ko:2016uft,Ko:2016fcd,Foot:2016wvj,Chacko:2016kgg,Foot:2017dgx,Foot:2018dhy,Buen-Abad:2022kgf,Joseph:2022jsf,Bansal:2022qbi,Klypin:1999uc,1999ApJ...524L..19M,2009ApJ...700.1779Z,2011ApJ...739...38P,Klypin:2014ira,2011MNRAS.416.3017B,Papastergis:2014aba,2010AdAst2010E...5D,2011AJ....141..193O,Kamionkowski:2022pkx,Philcox:2021kcw,ACT:2025fju,Homma:2023ppu,Wright:2025xka,Oman:2015xda,Zentner:2022xux,Schoneberg:2021qvd}. 
     
    There are also strong theoretical motivations to consider dark sectors that generate DAOs. Frameworks like the mirror-world scenario \cite{Blinnikov:1982eh,Blinnikov:1983gh,Kolb:1985bf,Goldberg:1986nk,Khlopov:1989fj,Foot:1991bp,Hodges:1993yb,Berezhiani:1995am,Foot:1999hm,Foot:2000vy,Mohapatra:2000qx,Mohapatra:2001sx,Foot:2002iy,Berezhiani:2003wj,Berezhiani:2003xm,Foot:2003eq,Foot:2003jt,Foot:2004pa,Berezhiani:2005ek,Foot:2014mia,Mohapatra:2017lqw} or Twin Higgs models \cite{Chacko:2005pe,Chacko:2005vw,Craig:2015pha,Craig:2016lyx} designed to restore symmetry or reduce tuning in the Standard Model can generically include dark sectors with matter and gauge content similar to the Standard Model, naturally leading to DM-DR interactions \cite{Alonso-Alvarez:2023bat}.
    
    As a phenomenon that can result from many different underlying particle models of dark matter, DAOs have been studied through the lens of effective models. The Effective THeory of Structure (ETHOS) formalism \cite{Cyr-Racine:2015ihg} parameterizes DM and DR interaction rates by power-laws of redshift controlled by a few coefficients. This enables a simple framework to be employed to solve the Boltzmann equations for the evolution of perturbations in the dark sector without reference to specific models. This phenomenological approach was further streamlined in \cite{Bohr:2020yoe} to characterize the DAO transfer function directly by the scale and height of the first DAO peak, although only for a fully interacting dark sector.  

    Many different cosmological probes across a wide range of scales can be impacted by these complex dark sector dynamics. The cosmic microwave background (CMB) power spectrum, CMB lensing, baryon acoustic oscillations (BAO), galaxy clustering and abundances, Milky Way satellite galaxies, cosmic shear tomography, Lyman-$\alpha$ forest, and the 21-cm cosmological signal are just some of the probes that are or have the potential to be sensitive to deviations from $\lcdm$ due to dark acoustic oscillations. Many studies have been performed to constrain or forecast the DAO signature using these probes, both in the ETHOS formalism and for specific models~\cite{Berezhiani:2003wj,Cyr-Racine:2013fsa,Cyr-Racine:2015ihg,Chacko:2016kgg,Lovell:2017eec,Archidiacono:2017slj,Kurmus:2022guy,Chacko:2018vss,Bansal:2021dfh,Bansal:2022qbi,Joseph:2022jsf,Buen-Abad:2017gxg,Krall:2017xcw,Murgia:2017lwo,Murgia:2018now,Garny:2018byk,Bose:2018juc,Sameie:2018juk,Archidiacono:2019wdp,Bohr:2020yoe,Bohr:2021bdm,Munoz:2020mue,Schneider:2018xba,Tan:2024cek}. On large cosmological scales probed by the CMB, agreement with $\lcdm$ is excellent, severely constraining the fraction of dark matter that can undergo DAOs on these scales to $\lesssim5\%$~\cite{Cyr-Racine:2013fsa,Bansal:2022qbi}. Observations of the Lyman-$\alpha$ forest have been used to rule out weak DAOs for 100\% of DM that are degenerate with warm dark matter (WDM) out to smaller scales~\cite{Murgia:2018now,Archidiacono:2019wdp,Bohr:2020yoe}. 
    
    Measurements of structure on smaller scales can more tightly constrain the scale at which DAOs appear, corresponding to increasingly early times at which the dark matter kinetically decoupled from other species. It has been shown using simulations that the non-linear evolution of the matter power spectrum washes out DAOs on small scales, and they largely disappear by $z\approx 5$ \cite{Bohr:2020yoe}, resulting only in suppressed power at high $k$, similar to WDM. However, it has also been found that DAOs are better preserved in the halo mass function (HMF) than the matter power spectrum \cite{Leo:2017wxg,Bohr:2020yoe,Bohr:2021bdm}. The suppression of power at small scales leads to a deficit and modulation in the number of low-mass dark matter halos compared to CDM. It is therefore particularly interesting to investigate observables that are both at higher redshifts and directly sensitive to the HMF.
    
    The high-redshift ultraviolet luminosity function (UVLF) of galaxies is an observable which has recently been shown to be sensitive to structure on scales of roughly $k=1$ to $10$ $h$/Mpc~\cite{Sabti:2021unj}. It depends strongly on the halo mass function through the connection of galaxies to the dark matter halos in which they reside. The combination of sensitivity to small scales, high redshift origin, and dependence on the HMF make this observable well-suited to the task of probing the DAO signature. 

    Motivated by these considerations, in this paper we derive the first constraints on general dark matter models with DAOs using measurements of the UVLF of galaxies at redshifts $z=3-9$ by the Hubble Space Telescope (HST), James Webb Space Telescope (JWST), Subaru Telescope, and Canada-France-Hawaii Telescope (CFHT). The non-linear behavior mentioned previously, along with baryonic effects, make the prediction of this observable significantly more difficult than the relatively simple computation of the CMB power spectrum. We perform N-body simulations to calibrate a semi-analytic formalism for predicting the HMF given an initial linear matter power spectrum, and employ a highly flexible model of the halo-galaxy connection to predict the UV luminosity function while marginalizing over uncertainty in the details of galaxy and star formation at high redshift. 
    
    To make our results as generally applicable as possible to models producing DAOs, we present our constraints in the parameter space of a new, phenomenological model for the linear transfer function $T^{2}(k)\equiv P(k)/P(k)_{\lcdm}$ which we introduce. This model allows a sub-unity fraction of the dark matter to be interacting and accurately captures the essential features of the DAO signature without reference to any particular particle physics model. The parameters of this model can be directly connected with easily interpretable, calculable physical quantities for a given model that gives rise to DAOs. We make use of this correspondence to place a few benchmark aDM and IDM models in the space of our transfer function parameters and demonstrate how our constraints can be re-cast. 

    The UV luminosity function quantifies the number density of galaxies per unit magnitude in the UV.  If the halo occupation distribution is assumed to be unity, then it can be computed as:
    \begin{equation}
        \Phi_{\mathrm{UV}}\equiv\frac{\mathrm{d}n_{\mathrm{h}}}{\mathrm{d}M_{\mathrm{UV}}} = \frac{\mathrm{d}n_{\mathrm{h}}}{\mathrm{d}M_{\mathrm{h}}}\times \frac{\mathrm{d}M_{\mathrm{h}}}{\mathrm{d}M_{\mathrm{UV}}} \,.
    \end{equation}
    The first factor is the halo mass function. We use the Extended Press-Schechter formalism to predict the HMF from the linear matter power spectrum, which we model with our phenomenological transfer function~\cite{1974ApJ...187..425P,1991ApJ...379..440B,Sheth:1999su,2002MNRAS.329...61S}. 
    
    The second factor is the halo-galaxy connection, which relates the mass of a host halo to the absolute magnitude $M_{\mathrm{UV}}$ of the  galaxy it contains. This function is of course dependent on the astrophysics of star formation, and can be generalized to include satellites or a sub-unity occupation fraction~\cite{Berlind:2001xk,Munoz:2023cup,2025A&A...699A.231S,2025arXiv250111674P}. We use the publicly available \texttt{GALLUMI} code to model the halo-galaxy connection \cite{Sabti:2021xvh}. \texttt{GALLUMI} is a pipeline for computing the high-redshift UVLF semi-analytically, implemented as a likelihood code for the cosmological Markov Chain Monte Carlo (MCMC) sampler \texttt{Monte Python}.
    With this pipeline, we can predict the UVLF as a function of the DAO transfer function parameters along with astrophysical nuisance parameters. 

    The paper is structured as follows. In Section \ref{sec:MPS} we introduce our new parameterization of the modified matter power spectrum. In Section \ref{sec:EPS} we discuss the Extended Press-Schechter formalism used to compute the halo mass function from the linear matter power spectrum, and in Section \ref{sec:simulations} the N-body simulations that were performed in order to calibrate it. Section \ref{sec:GALLUMI} describes the halo-galaxy connection model and computation of the UVLF using the code \texttt{GALLUMI} \cite{Sabti:2021xvh}. We show the results of our MCMC scans in Section \ref{sec:Results}, and conclude in Section \ref{sec:Conclusion}. 
 
    \section{Transfer Function Model}
    \label{sec:MPS}

    The first step in determining the constraints on a modified matter power spectrum is computing the primordial linear matter power spectrum. For $\lcdm$ as well as many models of interactions in the dark sector, this can be accomplished through the use of a Boltzmann-solving code like \texttt{CLASS} \cite{Lesgourgues:2015wza,Archidiacono:2017slj,Archidiacono:2019wdp,Bansal:2022qbi}. However, computing the matter power spectrum in models with DAOs up to the high $k$ required to predict observables like the high-redshift UVLF can be computationally expensive. This prohibits the performance of MCMC scans that would usually be conducted to obtain parameter estimates. Furthermore, many models which are distinct at the particle physics level result in very similar phenomenology at the level of the matter power spectrum. It is therefore desirable to find a model-agnostic parameterization of these common features. Significant progress was made in this direction for models of 100\% interacting dark matter within the ETHOS framework in \cite{Bohr:2020yoe}. They used a two-parameter model for the scale $\kpeak$ and height $\hpeak$ of the first DAO peak in the transfer function, with a fixed amount of damping for subsequent oscillations. The $\kpeak$ parameter was found to be strongly correlated with the time of DM decoupling from the DR, and $\hpeak$ was found to be strongly correlated with the ETHOS parameter $n$, equal to the ratio of the DM drag opacity to Hubble rate at the time of decoupling. Unfortunately, this parameterization is unable to accommodate a fractional abundance of interacting dark matter, which we are motivated to consider by existing constraints which limit the fraction of dark matter that can be interacting to a few percent in some parts of parameter space \cite{Bansal:2022qbi,Roy:2023zar,Roy:2024bcu}. 
    
\begin{figure}
        \centering
        \includegraphics[trim={0cm 0cm 0cm 0cm},clip,width=0.8\textwidth]{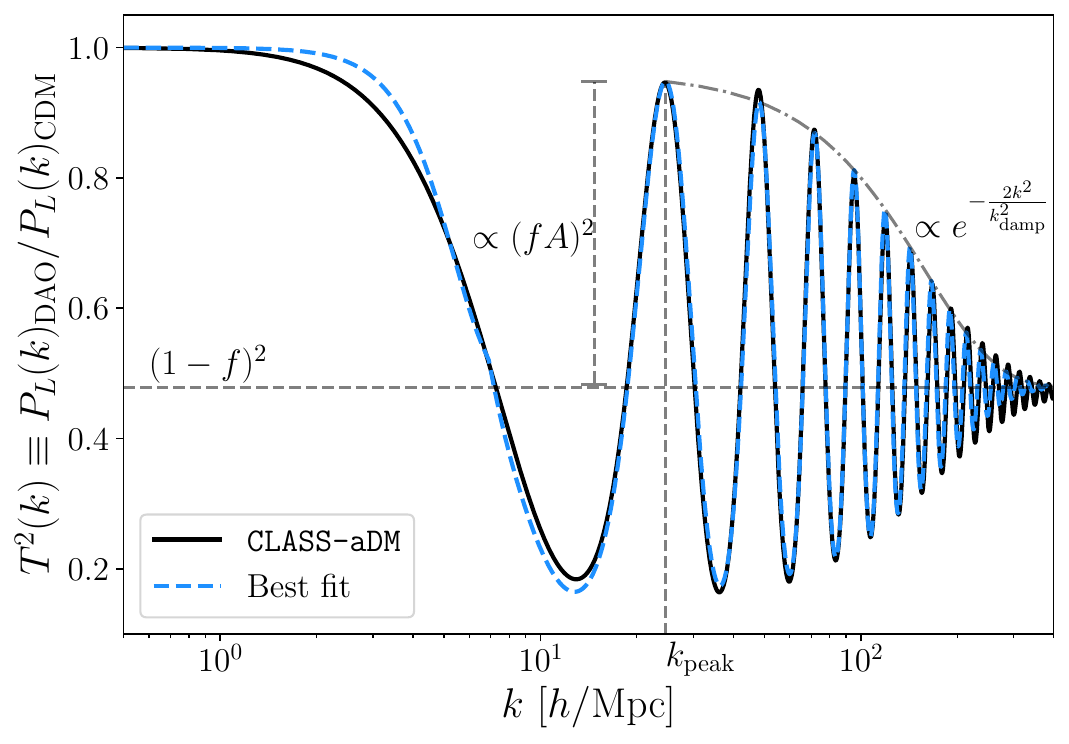}
        \caption{An example transfer function obtained using $\texttt{CLASS-aDM}$ and the best-fit $T(k)$ model, along with illustrations of the effect of each model parameter on the transfer function shape. Because we are interested in probing oscillatory behavior in the matter power spectrum, we prioritize matching the DAOs with our model over exactly matching the initial suppression of the power spectrum. The discrepancy between the best-fit model and the true transfer function at low $k$ is at the few-percent level, and extremely subdominant to the large deviation from a $\lcdm$ power spectrum that drives our constraints. The atomic dark matter parameters used to generate this transfer function are $\fd=0.3$, $\xid=0.1$, $\mpd=m_{p}$, $\med=m_{e}$, $\alphad=3\alpha$, using the notation of \cite{Bansal:2022qbi}.}
        \label{figure:adm_tk}
    \end{figure}

    We therefore introduce a new phenomenological model of the modified linear matter power spectrum relative to $\lcdm$, so that $P(k,z) = T^{2}(k)P_{\lcdm}(k,z)$. $T(k)$ is a simple function with four free parameters that captures the DAO signature of suppression and damped oscillations at small scales. The model has several fixed parameters which are calibrated by matching power spectra in the atomic dark matter model, computed with a Boltzmann code. Atomic dark matter is a simple, well-studied model where interactions of two new stable fermions with asymmetric abundances and opposite charges under a dark electromagnetism lead to strong DAOs \cite{Kaplan:2009de,Kaplan:2011yj,Cline:2012is,Cyr-Racine:2012tfp}. We use the model in this work as a physical point of comparison for the phenomenological model we develop, in order to demonstrate that it produces realistic power spectra with DAOs. The atomic dark matter power spectra were computed with the modified \texttt{CLASS} version \texttt{CLASS-aDM} \footnote{\texttt{https://github.com/jp-barron/class\_adm-3.1.git}}. Figure \ref{figure:adm_tk} illustrates the features of the model and the effect of each parameter on the shape of the transfer function by showing an example atomic dark matter transfer function along with the best-fit model.
    
    We model the transfer function as a sum of two parts, $T=T_{\alpha\beta\gamma\delta} + T_{\mathrm{osc}}$. $T_{\alpha\beta\gamma\delta}$ accounts for the suppression of the matter power spectrum by a mixed WDM/CDM-like function, identical to the $\alpha \beta \gamma \delta$ model of \cite{Hooper:2022byl}, with the replacement $f\equiv 1-\delta$: 
    \begin{equation}
    T_{\alpha \beta \gamma \delta}(k) \equiv f(1 + (\alpha k)^{\beta})^{\gamma} + (1-f) \,.
    \label{model:fabgd}
    \end{equation}
    This function transitions smoothly between 1 and $1-f$ as $k$ increases, with the transition occurring at a scale controlled by $\alpha$, defined below. The parameters $\beta=4.15$ and $\gamma=-20$ are fixed, while the depth of the suppression $f$ is free. 
    
    The DAOs are captured by $T_{\mathrm{osc}}$, modeled as sinusoidal oscillations which have a maximum at a wave number $\kpeak$ and have a fixed angular frequency $\omega=2.083\pi$, determined by fitting to aDM power spectra. The oscillations are turned on at wave number $k_{\mathrm{start}}=(1-\frac{3}{4}\frac{2\pi}{\omega})\kpeak$ with a Heaviside step function, at the start of the down-going oscillation before $\kpeak$. The oscillations have a Gaussian damping envelope with characteristic scale $\kdamp$, and an amplitude proportional both to $f$ as well as a parameter $A$. In summary, 
    \begin{equation}
    T_{\mathrm{osc}}(k) \equiv \Theta(k -k_{\mathrm{start}}) \left[fA \cos\left(\omega \left(\frac{k}{\kpeak} - 1\right)\right)e^{-\left(\frac{k}{\kdamp}\right)^{2}}\right] \,.
    \label{model:osc}
    \end{equation}
    The parameter $\alpha$ in Eq.~\eqref{model:fabgd} is defined in terms of $\kpeak$ so that 
    \begin{equation}
    \alpha = \frac{1}{k_{\mathrm{start}}}\left((0.1)^{1/\gamma}-1\right)^{1/\beta} \,.
    \end{equation}
    This ensures that $T_{\alpha\beta\gamma\delta}(k_{\mathrm{start}}) = 0.1 f + (1-f)$, so that the suppression of the power spectrum by $T_{\alpha\beta\gamma\delta}$ begins at the correct scale to match onto $T_{\mathrm{osc}}$. The free parameters of the model are $f$, $\kpeak$, $\kdamp$, and $A$.
 
    \begin{figure}
        \centering
        \includegraphics[trim={0cm 0cm 0cm 0cm},clip,width=0.8\textwidth]{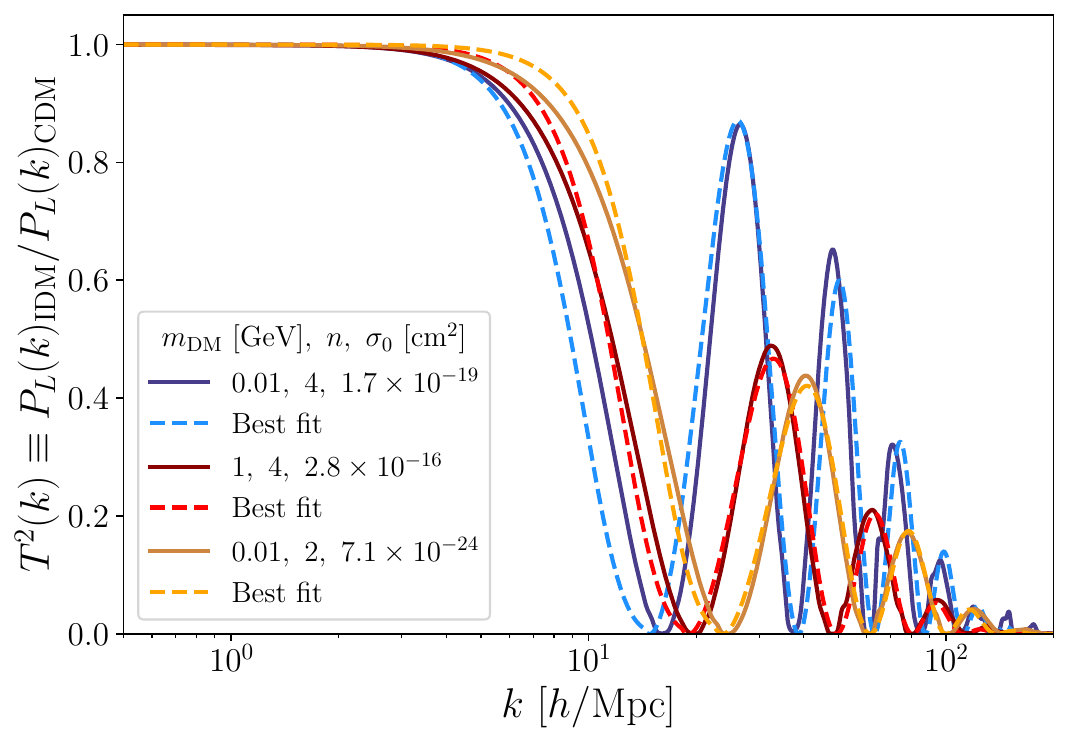}
        \caption{Three transfer functions generated for a model with DM-baryon elastic scattering interactions using a modified version of CLASS \cite{Gluscevic:2017ywp,Boddy:2018kfv,Nguyen:2021cnb}, along with the best fits from our model. These parameter choices are taken from \cite{Nadler:2025fcv}, in which single-halo zoom-in simulations were performed for alternative dark matter models. $n$ is the index of the power law governing the DM-baryon scattering rate. We successfully match the numerically computed transfer functions to within a few percent in amplitude and phase, sufficient for the purpose of modeling the UVLF and comparing to data with much larger uncertainties.
        \label{figure:dmb_tk}}
    \end{figure}

\begin{figure}[htp]

\centering
\includegraphics[width=.33\textwidth]{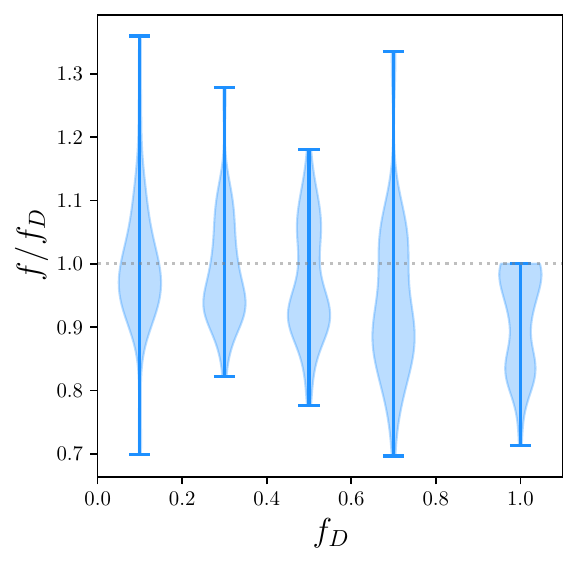}\hfill
\includegraphics[width=.33\textwidth]{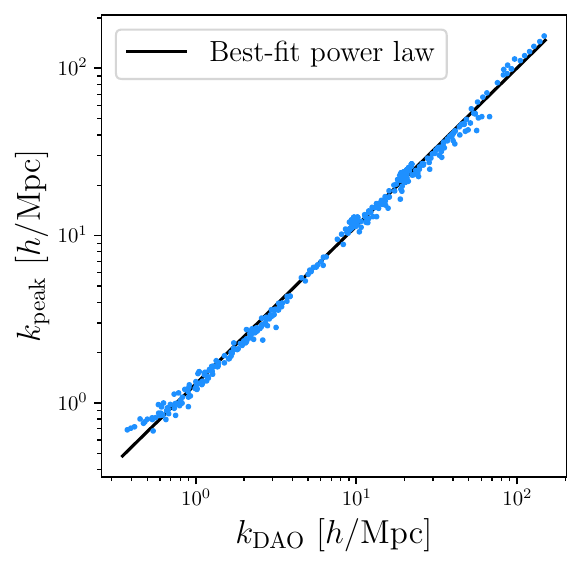}\hfill
\includegraphics[width=.33\textwidth]{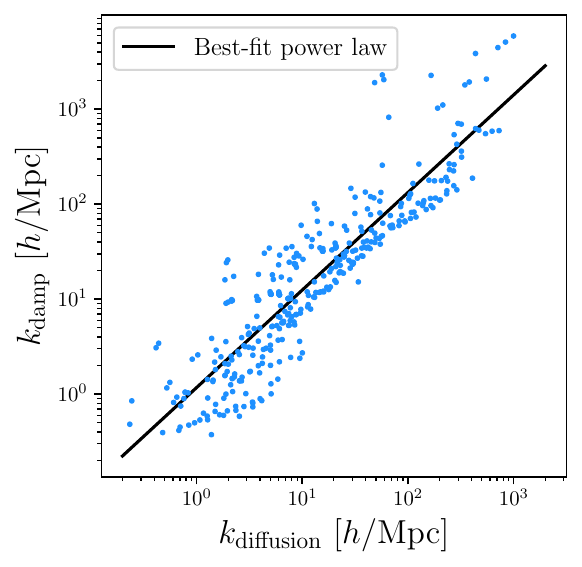}
\caption{Relationships between physical quantities and best-fit transfer function parameters for 380 different choices of aDM parameters. Left: Violin plots of the distribution of $f/\fd$ for $\fd\in \{0.1,\ 0.3,\ 0.5,\ 0.7,\ 1\}$. The extremal values of $f/\fd$ are marked by horizontal lines. Center: DAO scale, with best-fit power law $\kpeak = 1.30 \left(\frac{\kdao}{h/\mathrm{Mpc}}  \right)^{0.944}$. Right: Dark diffusion damping scale, with best-fit power law $\kdamp=1.17 \left(\frac{k_{\mathrm{diffusion}}}{h/\mathrm{Mpc}}  \right)^{1.03}$.}
\label{fig:Tk_param_correlations}
\end{figure}

    To check that this model accurately captures the behavior of physical power spectra with DAOs, we compute power spectra for 380 points across the atomic dark matter parameter space and fit the model to the results. We also demonstrate the model's ability to fit power spectra from the IDM model with a few examples in Figure \ref{figure:dmb_tk}. We find that the best-fit model parameters are strongly correlated with easily understood physical parameters and scales. These relationships between physical quantities and best-fit model parameters are displayed in Figure \ref{fig:Tk_param_correlations}. First, we find that the parameter $f$ is highly correlated with the fraction of dark matter that is interacting, as expected. Consistent with the findings in \cite{Bohr:2020yoe}, $\kpeak$ is strongly correlated with the DAO scale $\kdao$, defined as $\kdao = 2\pi/r_{\rm{DAO}}$, where $r_{\rm{DAO}}$ is the dark sector sound horizon at the dark drag epoch. The damping scale $\kdamp$ is correlated with the diffusion damping scale $k_{\mathrm{diffusion}}$, consistent with the Gaussian form of diffusion damping \cite{Hu:1995en}. This correlation was not observed as strongly in the ETHOS work. This is likely due to the fact that atomic dark matter generically has a narrow DR visibility function due to the exponential decrease in the DM-DR interaction rate at the time of dark recombination. In contrast, low-$n$ ETHOS models have a much broader visibility function, resulting in the DM spending more time in the weakly coupled regime and changing the damping scale, although maintaining its Gaussian form. Best-fit power-law relations between $\kdao$ and $\kpeak$ as well as $k_{\mathrm{diffusion}}$ and $\kdamp$ are shown in Figure \ref{fig:Tk_param_correlations} to illustrate the correspondence for these parameters. Clearly there is much more scatter in the relation for $\kdamp$ than for $\kpeak$, limiting the ability to translate $\kdamp$ into $k_{\mathrm{diffusion}}$ one-to-one. The fraction of dark matter that is interacting, the dark sound horizon, and diffusion damping scale are all completely general quantities, enabling constraints on the phenomenological transfer function parameters to be reinterpreted in the context of specific models. 

    \section{Extended Press-Schechter Formalism}
    \label{sec:EPS}
    In order to predict the UVLF for arbitrary DAO transfer function parameters and compare to data, we must be able to quickly compute the halo mass function, without running new N-body simulations for every parameter point. The halo mass function $\frac{\mathrm{d}n_{\mathrm{h}}}{\mathrm{d}M_{\mathrm{h}}}$ is simply the number density of dark matter halos per mass interval. 
    
    It has been demonstrated both for CDM \cite{Jenkins:2000bv,Reed:2006rw} and alternative dark matter models \cite{Schneider:2013ria,Sameie:2018juk,Bohr:2021bdm} that the halo mass function may be estimated semi-analytically from the linear matter power spectrum, through the extended Press-Schechter formalism \cite{1974ApJ...187..425P,1991ApJ...379..440B,Sheth:1999su,2002MNRAS.329...61S}. This formalism makes use of the statistics of first-crossing distributions in order to predict the fraction of matter that has collapsed into halos of a given mass at a particular time. The widely used Sheth-Tormen (ST) HMF \cite{Sheth:1999su,2002MNRAS.329...61S}, based on an ellipsoidal collapse model, has been calibrated and validated against simulations \cite{Reed:2006rw} for CDM. It is conventionally defined as follows: 
    \begin{equation}
        \frac{\mathrm{d}n_{\mathrm{h}}}{\mathrm{d}M_{\mathrm{h}}} = \frac{\bar{\rho}_{\mathrm{m}}}{M_{\mathrm{h}}}\frac{\mathrm{d}\ln{\sigma^{-1}_{M_{\mathrm{h}}}}}{\mathrm{d}M_{\mathrm{h}}}f_{\rm{ST}}(\sigma_{M_{\mathrm{h}}}) \,.
    \end{equation}
    Here, $\bar{\rho}_{\mathrm{m}}$ is the average matter density and $\sigma^{2}_{M_{\mathrm{h}}}$ is the variance of density perturbations smoothed on a scale corresponding to $M_{\mathrm{h}}$.
    The Sheth-Tormen first-crossing distribution is defined as 

    \begin{equation}
        f_{ST}(\nu) = A\sqrt{\frac{2q\nu}{\pi}}(1 + (q\nu)^{-p})\exp \left(-\frac{q\nu}{2} \right) \,,
        \label{eq:ST_HMF}
    \end{equation}
    where $\nu \equiv \delta_{ST}^{2}/\sigma^{2}_{M_{\mathrm{h}}}$ and $\delta_{ST}=1.686$. The parameter $q$ is calibrated to simulation, $p=0.3$, and $A=0.3222$ is determined by requiring that $f_{ST}$ is normalized correctly. The variance of perturbations $\sigma^{2}_{M_{\mathrm{h}}}$ is defined in equation~\eqref{eq:sigma2} by integrating the linear power spectrum with a window function $W_{M_{\rm{h}}}(k)$: 
    \begin{equation}
        \sigma^{2}_{M_{\mathrm{h}}} = \int \frac{d^{3}k}{(2\pi)^{3}}W_{M_{\rm{h}}}^{2}(k)P(k,z) \,.
        \label{eq:sigma2}
    \end{equation}
    The commonly used window function for CDM is a real-space top-hat, which intuitively connects a distance scale $R$ with a halo mass $M_{\rm{h}}(R) = 4\pi \bar{\rho}_{m} R^{3}/3$. However, for models of dark matter that predict a truncated matter power spectrum, such as warm dark matter, it has been found that the real-space top-hat window function fails, and a step function in Fourier space, the so-called ``sharp-$k$" window function, agrees better with the halo mass function obtained from simulation \cite{2013MNRAS.428.1774B,Schneider:2013ria}. A further change was proposed in \cite{Leo:2018odn} to use a smoothed version of the sharp-$k$ window function, which is more accurate at low halo masses for sharply truncated power spectra. This window function has two free parameters: $\beta$, which controls the smoothness of the step, and $c$, which controls the correspondence between $R$ and $M_{\rm{h}}$. It is defined by: 
    \begin{equation}
    W_{M_{\rm{h}}}(k) = W(k;R(M_{\mathrm{h}})) = \frac{1}{1 + \left(\frac{kR}{c}\right)^{\beta}} \,.
    \end{equation}   
    The smooth-$k$ window function was employed in \cite{Sameie:2018juk,Bohr:2021bdm} in order to study the halo mass function of ETHOS models of interacting dark matter and dark radiation. The EPS parameters were calibrated against DM-only N-body simulations with ETHOS power spectra that were conducted and described in \cite{Bohr:2020yoe}, using their two-parameter model for the transfer function. Because our model includes the possibility of sub-unity fractions of interacting dark matter, we must validate or re-calibrate the free parameters $q, \beta$, and $c$, necessitating new N-body simulations.

    \section{Simulations}
    \label{sec:simulations}
    Similar to \cite{Bohr:2020yoe}, we perform DM-only N-body simulations using a zoom-in technique to economically resolve the non-linear matter power spectrum and halo mass function across the wide range of scales necessary for comparison to the high-redshift UVLF. Specific models could have additional self-interactions or dissipation within the dark sector, but we do not include these here; even for models with such processes, there are typically parts of parameter space where these processes are unimportant, particularly at the high redshifts we are sensitive to. As a specific example, for atomic dark matter, dissipation can be avoided if the dark hydrogen remains bound at the virial temperature of the dark matter halo \cite{Roy:2024bcu}. These simulations also do not include baryonic effects, because we wish to isolate the impact of the dark matter physics relative to CDM, and because we use a phenomenological model of the halo-galaxy connection rather than obtaining star formation rates from the simulations.  
    
    The \texttt{GIZMO} multi-physics code \cite{Hopkins:2014qka,2005MNRAS.364.1105S} is used to perform the simulations, with initial conditions generated using the \texttt{MUSIC} code \cite{2011MNRAS.415.2101H}. The gravity solver in \texttt{GIZMO} is a hybrid tree-particle mesh method. Simulations are run for interacting dark matter fractions of 0.1, 0.5, and 1, and $\kdao$ of 5, 20, 66, and 190 $h$/Mpc, as well as CDM, for a total of thirteen simulations. These choices span the parameter space to which the high-redshift UVLF could reasonably be sensitive. Linear power spectra with these properties are generated with the $\classadm$ code at $z=99$ and used as the input to the initial conditions code. The $\lcdm$ parameters are set to Planck 2018 best-fit values \cite{Planck:2018vyg}. The initial low-resolution simulation volume is $(40\ \mathrm{Mpc}/h)^{3}$, with $N=512^{3}$ particles. The comoving softening length is 6.814 kpc/$h$. After evolving the simulation from $z=99$ down to $z=4$, the (non-linear) matter power spectrum is computed. Sub-regions of this box of size $(6.25\ 
    \mathrm{Mpc}/h)^{3}$ are then examined to find a sub-region with close to average density and a matter power spectrum most similar to that of the whole box. The figure of merit $L$ used to identify the chosen sub-region is an integral of the relative difference between the dimensionless matter power spectra of the sub-region and the full box, on the range of scales $k=5$ to $20$ $h$/Mpc at $z=5$:
    \begin{equation}
        L = \int_{5\ h/\mathrm{Mpc}}^{20\ h/\mathrm{Mpc}} dk \bigg\lvert\frac{\Delta^{2}(k)_{\rm sub} - \Delta^{2}(k)_{\rm full}}{\Delta^{2}(k)_{\rm full}}\bigg\rvert \,.
        \label{eq:Pksim}
    \end{equation} 
    The sub-region with minimal $L$ and relative overdensity of the sub-region $|\delta_{\rm sub}| < 0.05$ is chosen, then expanded to an $(8\ \mathrm{Mpc}/h)^{3}$ box in order to encompass the Lagrangian volume of particles that end up within the region of interest. The simulation is then re-run as a zoom-in, with this sub-region populated at an effective resolution of $N_{\rm{eff}}=4096^{3}$ particles. The gravitational softening length in the zoom-in region is 0.125 comoving kpc/$h$. 
    
    The halo mass function for each simulation is extracted with the aid of the halo-finder code \texttt{Rockstar} \cite{2013ApJ...762..109B}, which is run on both the big box and zoom-in simulations. The distribution of halo masses is then binned from $10^{7}-10^{14}$ $M_{\odot}/h$ to compute the HMF. The HMF from the big box is used for halo masses above $M_{h} = 2\times 10^{9} M_{\odot}/h$, below which the zoom-in HMF is used. A finite-volume correction is applied to the high-halo mass part of the HMF as in \cite{Bohr:2021bdm}, accounting for the fact that modes larger than the periodic simulation box length cannot evolve by setting $\sigma^{2}(M)\rightarrow \sigma^{2}(M)-\sigma^{2}(M_{\mathrm{box}})$. 

    \begin{figure}
    \centering
    \includegraphics[width=0.8\textwidth]{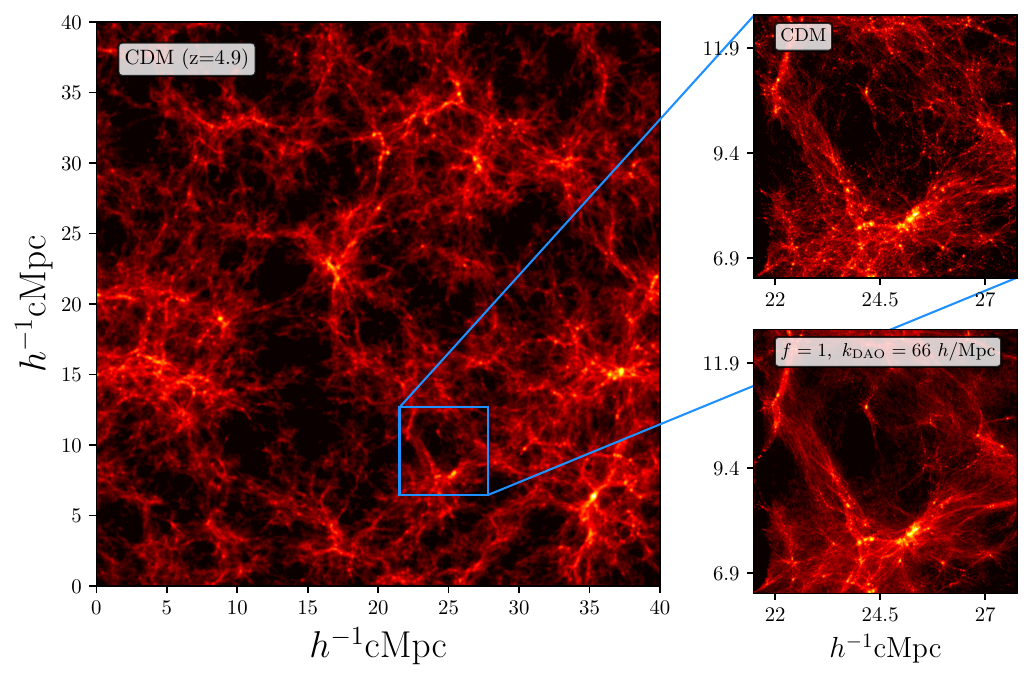}
    \caption{Images of the matter density in our simulations at $z=4.9$. The low-resolution CDM simulation is on the left, with the chosen zoom-in region highlighted. On the right are the high-resolution zoom-in regions for both CDM and the DAO model with $f=1$, $\kdao=66\ h/\mathrm{Mpc}$. The smoothing of structure on small scales is clearly visible.
    \label{figure:simulation}}
    \end{figure}    
    
    \begin{figure}
    \centering
    \includegraphics[width=0.45\textwidth]{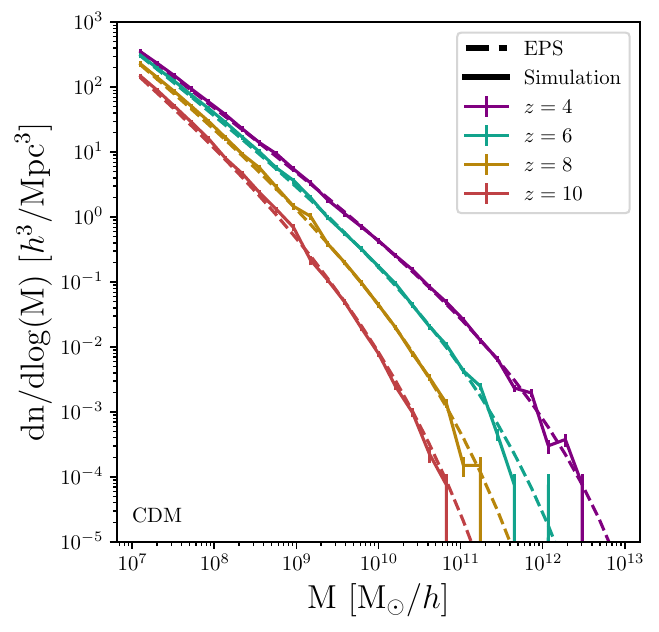}
    \includegraphics[width=0.45\textwidth]{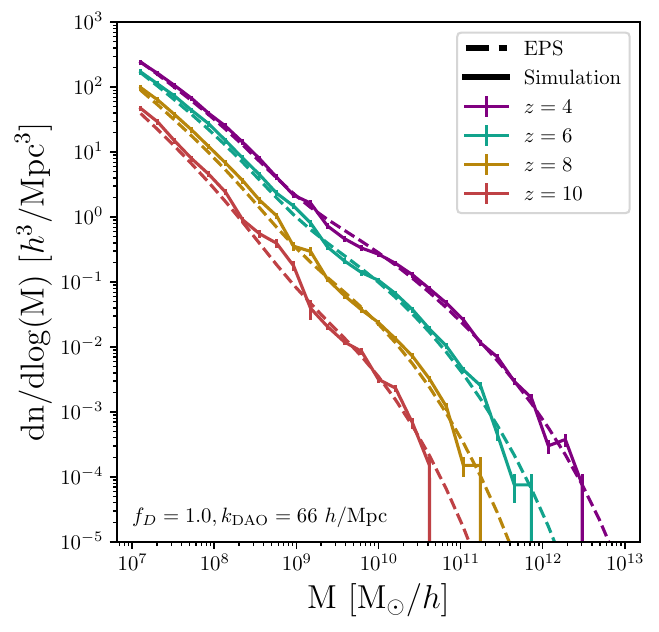}
    \caption{Halo mass function derived from our simulations (solid) and analytic theory (EPS, dashed) for CDM (left) for aDM with $f = 1$, $\kdao \approx$ 66 $h$/Mpc (right) at various redshifts relevant to UVLFs.
    \label{figure:hmf}}
    \end{figure}
    
    Once an HMF is obtained for each simulation, a least-squares regression fit for the EPS window-function parameters $(c,\beta,q)$ is performed, comparing the semi-analytic HMFs to the ones obtained from all the simulations at redshifts from 4 to 12. The $\chi^{2}$ to be minimized is 
    \begin{equation}
        \chi^2(c, \beta, q) = \sum_j \sum_i \frac{\left[\frac{\rmd n_{h, \rm sim}}{\rmd \log M_h} (M_{h,i},z_j) - \frac{\rmd n_{h, \rm EPS}}{\rmd \log M_h} (M_{h,i}, z_j) \right]^2}{\sigma_{\rm stat}^2(M_{h,i}, z_j) + \sigma_{\rm sys}^2 (M_{h,i}, z_j)} \,.
    \end{equation}
   
    The $\chi^{2}$ is summed over nine redshift snapshots $z_j = 4, \cdots, 12$, and 30 mass bins $M_{h,i}$. The statistical uncertainty $\sigma_{\mathrm{stat}}(M_{h,i}, z_j)$ is the Poisson uncertainty in the number of halos identified in the simulation box for a given mass bin $M_{h,i}$ at redshift $z_j$. A uniform relative systematic uncertainty $\sigma_{\mathrm{sys}}$ for every bin in each redshift is also imposed in order to account for any other sources of uncertainty in the simulation. 
    Because the inherent systematic uncertainty in the simulations is unknown, the fit is performed for benchmark values of $\sigma_{\mathrm{sys}}$ of 3\%, 10\%, and 30\%.
    The best-fit parameters have some dependence on the choice of systematic uncertainty. For $\sigma_{sys}=3\%$, the best-fit values are $(c,\beta,q)$=(3.11, 2.548, 0.904). For $\sigma_{sys}=10\%$, the best-fit values are (2.978, 2.516, 0.900). For $\sigma_{sys}=30\%$, the best-fit values are (2.888, 2.646, 0.911). We perform MCMC scans comparing to UVLF data with each set of best-fit EPS parameters and find that the constraints are insensitive to this level of variation, so we report results only for the 10\% benchmark. Figure \ref{figure:hmf} shows the HMF extracted from simulation along with the best EPS fit for both CDM and one of the DAO parameter choices. The suppression and oscillation of the HMF at low halo mass is clearly visible, and the EPS model successfully reproduces the simulation-derived HMF across the redshifts of interest for both CDM and DAO power spectra. The same behavior has been observed in previous work that has studied the halo mass function for models with DAOs~\cite{Bohr:2020yoe,Bohr:2021bdm}. 

    With this calibration complete, the job of the N-body simulations is accomplished. We are now equipped to quickly generate a linear matter power spectrum with DAOs and map it to an HMF through the EPS formalism. The final ingredient to obtain the UVLF is the halo-galaxy connection. 
    
    \section{The Halo-Galaxy Connection}
    \label{sec:GALLUMI}

    We model the halo-galaxy connection, relating the mass of a host halo to the absolute magnitude $M_{\mathrm{UV}}$ of the galaxy it contains, with the \texttt{GALLUMI} code~\cite{Sabti:2021xvh}. \texttt{GALLUMI} is a pipeline for computing the high-redshift UVLF semi-analytically, implemented as a likelihood code for the cosmological MCMC sampler \texttt{Monte Python}~\cite{Brinckmann:2018cvx}.
    
    This code has three different models for the halo-galaxy connection, each making different assumptions about the relationship between halo mass and average galactic UV magnitude. The authors of \texttt{GALLUMI} found that the three models lead to consistent cosmological results. We perform MCMC scans with two of these, and find that the constraints we derive are insensitive to the choice. We present the first model here, which is also used in \cite{Sabti:2021unj} as their fiducial model, and similar to those used in~\cite{Wang:2023xmm,Wang:2024piu}. Appendix \ref{sec:appendix} describes the second model and the results obtained with it. 
    
    The model assumes that young, massive stars dominate the emission of UV light at high redshifts, so that the star formation rate of a galaxy determines its UV luminosity \cite{1983ApJ...272...54K,Salim:2007is}. The model also assumes that the galactic star formation rate (SFR) $\dot{M}_{*}$ is related to the halo accretion rate $\dot{M}_{\mathrm{h}}$ on average by a double-power law, whose normalization, pivot scale, and power law indices are allowed to float. This functional form is motivated by simulations and observations of galaxy formation \cite{Wechsler:2018pic,2016MNRAS.460..417S,2019MNRAS.488.3143B} and in particular is similar to the \textsc{Emerge} model introduced in \cite{2018MNRAS.477.1822M}, 

    \begin{equation}
        \tilde{f}_{*} = \frac{\dot{M}_{*}}{\dot{M}_{\mathrm{h}}} = \frac{\epsilon_{*}}{\left( \frac{M_\mathrm{h}}{M_{c}}\right)^{\alpha_{*}} +\left( \frac{M_\mathrm{h}}{M_{c}}\right)^{\beta_{*}} } \,,
    \end{equation}
    where $\alpha_{*} \leq 0$, $\beta_{*} \geq 0$, $\epsilon_{*} \geq 0$ and $M_{c}\geq 0$ are free parameters. Both $\alpha_{*}$ and $\beta_{*}$ are independent of redshift, while $\epsilon_{*}$ and $M_{c}$ are allowed to evolve as power-laws of redshift:

    \begin{equation}
        \log_{10}\epsilon_{*}(z) = \epsilon_{*}^{\mathrm{s}}\times \log_{10}\left(\frac{1+z}{1+6}\right) + \epsilon_{*}^{\mathrm{i}} \,,
    \end{equation}
    \begin{equation}
        \log_{10}\frac{M_{c}(z)}{M_{\odot}} = M_{c}^{\mathrm{s}} \times \log_{10}\left(\frac{1+z}{1+6}\right)+M_{c}^{\mathrm{i}} \,.
    \end{equation}
    
    The halo accretion rate is assumed to be given by a semi-analytical description derived from the EPS formalism, which yields exponential growth of halo mass during matter domination followed by a power law during dark energy domination \cite{Correa:2014xma},
    \begin{equation}
        \dot{M}_{\mathrm{h}} = -\sqrt{\frac{2}{\pi}} \frac{(1+z)H(z)M_{\mathrm{h}}}{\sqrt{\sigma^{2}_{M_{\mathrm{h}}}(Q) - \sigma^{2}_{M_{\mathrm{h}}}}} \frac{1.686}{D^{2}(z)} \frac{\mathrm{d}D(z)}{\mathrm{d}z} \,.
    \end{equation}
    The accretion rate model has one free parameter $Q$, related to the minimum main progenitor mass in the halo merger tree algorithm, which is allowed to vary from 1.5 to 2.5, consistent with previous work \cite{Neistein:2006ak,Correa:2014xma,Schneider:2020xmf}. The SFR is then converted to a UV luminosity by a simple factor \cite{Madau:1997pg,Kennicutt:1998zb,Madau:2014bja}:
    
    \begin{equation}
    \dot{M}_{*} = \kappa_{\mathrm{UV}}L_{\mathrm{UV}} \,,
    \label{eq:SFR-LUV}
    \end{equation}
    where $\kappa_{\mathrm{UV}}=1.15\times 10^{-28} M_{\odot}\ \mathrm{s}\ \mathrm{erg}^{-1}\ \mathrm{yr}^{-1}$ which is degenerate with the normalization $\epsilon_\star$ of the SFR model and thus we simply keep fixed.
    The absolute magnitude is then related to the UV luminosity by 
    \begin{equation}
        \log_{10}\left(\frac{L_{\mathrm{UV}}}{\mathrm{erg}\ \mathrm{s}^{-1}} \right) = 0.4 \times (51.63 - M_{\mathrm{UV}})
    \end{equation}
    Finally, stochasticity in the relation between halo mass and UV luminosity is accounted for by assuming a normal distribution for $M_{\mathrm{UV}}$ with mean computed according to the above procedure, and standard deviation $\sigma_{M_{\mathrm{UV}}}$ as a free parameter.

    \begin{equation}
        P(M_{\mathrm{UV}}) = \frac{1}{\sqrt{2\pi}\sigma_{M_{\mathrm{UV}}}}\exp\left[-\frac{(M_{\mathrm{UV}} - \langle M_{\mathrm{UV}} \rangle)^{2}}{2\sigma_{M_{\mathrm{UV}}}^{2}}\right] \,.
    \end{equation}
  
    \begin{figure}
    \centering
    \includegraphics[trim={0cm 0cm 0cm 0cm},clip,width=0.8\textwidth]{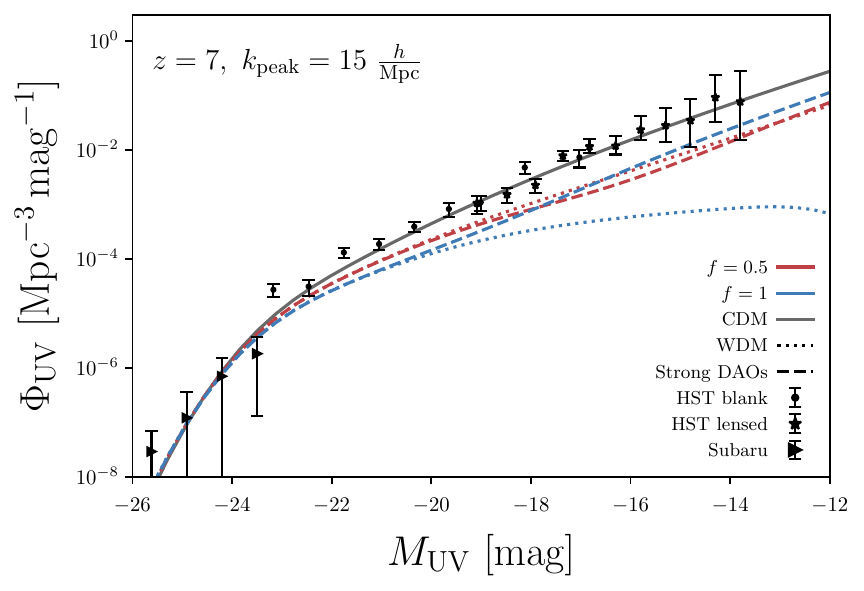}
    \caption{The UVLF at redshift 7 for CDM, strong DAOs, and WDM-like transfer functions. The strong DAO and WDM-like model have $\kpeak=15\ h/\mathrm{Mpc}$, and are plotted for both $f=0.5$ and $f=1$. The strong DAO model has $A=1$ and $\kdamp/\kpeak=10$, while the WDM model has $A=0$. All UVLF predictions have the same set of astrophysical nuisance parameters. For $f=1$, the faint end of the UVLF is more suppressed with $A=0$ than $A=1$. However, for $f=0.5$, the UVLF is slightly more suppressed with $A=1$ around $M_{\mathrm{UV}}=-16$ before rising again. Data from~\cite{2021AJ....162...47B,2022ApJ...940...55B,2022ApJS..259...20H} are shown with error bars. 
    \label{figure:UVLF_modelcomparison}}
    \end{figure}  
     
    This model of star formation rate is highly flexible, with eight parameters that are marginalized over to account for the large uncertainty in star formation at early times:  $\alpha_{*},\ \beta_{*},\ \epsilon_{*}^{\mathrm{s}},\ \epsilon_{*}^{\mathrm{i}},\ M_{c}^{\mathrm{s}},\ M_{c}^{\mathrm{i}},\ Q,\ \sigma_{M_{\mathrm{UV}}}$. If there are any effects on the galaxy-halo connection independent of the halo mass function, as long as they can be captured by the functional form of the SFR in \texttt{GALLUMI}, we marginalize over them. We leave a more detailed investigation of the effects of non-cold DM on star formation rates of individual halos to future work.

    To demonstrate the impact of the DAOs on the UVLF, we compare the predicted UVLF for CDM with four different DAO scenarios in Figure \ref{figure:UVLF_modelcomparison}, all with the same astrophysical parameters. Strong DAOs are compared with a mixed WDM/CDM-like transfer function for both $f=0.5$ and $f=1$. All four parameter choices have $\kpeak=15\ h/\mathrm{Mpc}$, corresponding roughly to $m_{\mathrm{WDM}}\approx 0.5$ keV. The strong DAOs have $A=1$ and $\kdamp=150\ h/\mathrm{Mpc}$, while the WDM-like model has $A=0$. In all four scenarios, the faint end of the UVLF is significantly suppressed relative to CDM. However, the absence or presence of DAOs relative to just a mixed WDM-like power spectrum significantly affects the UVLF across a large range of magnitudes, in an $f$-dependent way. For $f=1$, the presence of large-amplitude DAOs in the power spectrum drastically enhances the faint end of the UVLF relative to the $A=0$ case. This is because the WDM-like model has no power at small scales, so the DAOs only add power in comparison. In contrast, for $f=0.5$, the UVLF exhibits slightly more suppression at intermediate magnitudes with strong DAOs than with no DAOs. In this case, the WDM-like power spectrum does not fall to zero, and the trough of the first DAO leads to a relative suppression of the halo mass function and UVLF. 

    With the halo-galaxy connection model in place, we are now equipped to predict the UVLF starting from transfer function parameters, and to obtain constraints.    
    
    \section{Results}    \label{sec:Results}

    We obtain constraints by running Markov Chain Monte Carlo (MCMC) scans to estimate the posterior distribution of the model parameters given data. 
    
    The datasets we consider are measurements of the UVLF ranging over redshifts 3 to 9 from several different telescopes. We include data from the Hubble Space Telescope using both blank and lensed fields, the James Webb Space Telescope, the Subaru Telescope, and the Canada-France-Hawaii Telescope (CFHT)~\cite{2021AJ....162...47B,2022ApJ...940...55B,2024MNRAS.533.3222D,2022ApJS..259...20H}.
    Data from the HST blank-field survey are used for $z\sim 3,4,5,6,7,8$, while HST lensed-field data extend from $z=3-9$. 
    Data from Subaru and CFHT are used for $z\sim 4,5,6,7$. The Hubble lensed field data extend to fainter magnitudes than the blank field, while the Subaru/CFHT data extend to brighter magnitudes. For redshifts and magnitudes where both HST and Subaru/CFHT measurements are available, we use the Subaru/CFHT data to avoid double-counting. We use the JWST data for $z=9$, since it has smaller uncertainties than the HST blank-field data. At redshifts 10 and above, the UVLF as measured by JWST is unusually high, and challenging to reconcile with lower redshift measurements even in CDM~\cite{2023MNRAS.526.2665S,2023MNRAS.519..843M,Munoz:2023cup}. We therefore omit it in this work. 

    This combination of datasets constrains the behavior of the UVLF across a much wider range of magnitudes than the HST blank field data alone, which were previously used in \cite{Sabti:2021unj}. This aids in removing degeneracies with astrophysical nuisance parameters, enabling stronger constraints to be set on the DAO signature.

    We fix the $\lcdm$ parameters to their best-fit values from the Planck 2018 results~\cite{Planck:2018vyg}. In preliminary scans allowing the $\lcdm$ parameters to float, constrained by Planck data, they were found to have negligible effect on the constraints on the DAO model parameters.  
    We use the MCMC code \texttt{Monte Python} to run the scans, using the usual Metropolis-Hastings algorithm~\cite{Brinckmann:2018cvx}. The UVLF likelihood code computes the UVLF at the redshifts and magnitudes corresponding to the data, integrating the contributions from galaxies across each bin in magnitude. The data are assumed to have asymmetric Gaussian uncertainties. 
    
    The transfer function parameters are sampled with uniform priors on $f$, $A$, $\kpeak^{-1}$, and $\log_{10}(\kdamp/\kpeak)$. We choose a uniform prior on $\kpeak^{-1}$ in order to avoid the volume effects that would arise if $\kpeak$ or $\log_{10}(\kpeak)$ were sampled uniformly, due to the degeneracy of the $\kpeak\rightarrow \infty$ limit with $\lcdm$. The upper bound on $\kpeak^{-1}$, corresponding to $\kpeak > 2\ h/\mathrm{Mpc}$, is set to minimize overlap with the regime in which the CMB power spectrum would be significantly impacted by the DAOs. Large-scale cosmological observations have been found to impose strict constraints on the fraction of dark matter undergoing acoustic oscillations \cite{Cyr-Racine:2013fsa,Bansal:2022qbi}. We are interested in the extension of those constraints to smaller scales to which the UVLF is sensitive. The prior ranges for the DAO transfer function parameters and astrophysical nuisance parameters are given in Table \ref{tab:nuisance_priors}.

    \begin{table}
        \centering
        \begin{tabular}{ c c c }
        \hline
        Parameter & Minimum & Maximum \\
        \hline
        $f$ & 0 & 1\\
        $A$ & 0 & 1.2 \\
        $\kpeak^{-1}/(\mathrm{Mpc}/h)$ & 0 & 0.5\\
        $\log_{10}(\kdamp/\kpeak)$ & -0.3 & 1.7 \\
        \hline
        $\alpha_{*}$ & -3& 0 \\
        $\beta_{*}$ & 0& 3\\
        $\epsilon_{*}^{\mathrm{s}}$ & -3& 3\\
        $\epsilon_{*}^{\mathrm{i}}$ & -3& 3\\
        $M_{c}^{\mathrm{s}}$ & 0& 3\\
        $M_{c}^{\mathrm{i}}$ & 7& 15\\
        $\sigma_{M_{\mathrm{UV}}}$ & 0.001& 3\\
        $Q$& 1.5&2.5 \\
        $a_{\mathrm{ST}}$ & 0.9& 1\\
        \hline
        \end{tabular}
        \caption{Prior ranges for DAO and astrophysical nuisance parameters in \texttt{GALLUMI}.}
        \label{tab:nuisance_priors}
    \end{table}

    There are two limiting cases in the DAO parameter space that reduce to $\lcdm$ identically: $f\rightarrow 0$ and $\kpeak\rightarrow \infty$. The limit of $f\rightarrow 0$ corresponds to the fraction of dark matter that is not cold and collisionless becoming small, and therefore also the magnitude of its effects on structure formation. If $f$ is small enough for a given observable, the effects of the DAOs may be subdominant to the uncertainty in the data even for small $\kpeak$, generally corresponding to stronger dark sector interactions, while remaining consistent with $\lcdm$. 
    The limit of $\kpeak \rightarrow \infty$ leads to DAOs only on very small scales. For a given observable, sensitive to a limited range of scales, $\kpeak$ can be sufficiently large that even for $f=1$ the DAOs have no effect. 
    
\begin{figure}
    \centering
    \includegraphics[trim={0cm 0cm 0cm 0cm},clip,width=0.8\textwidth]{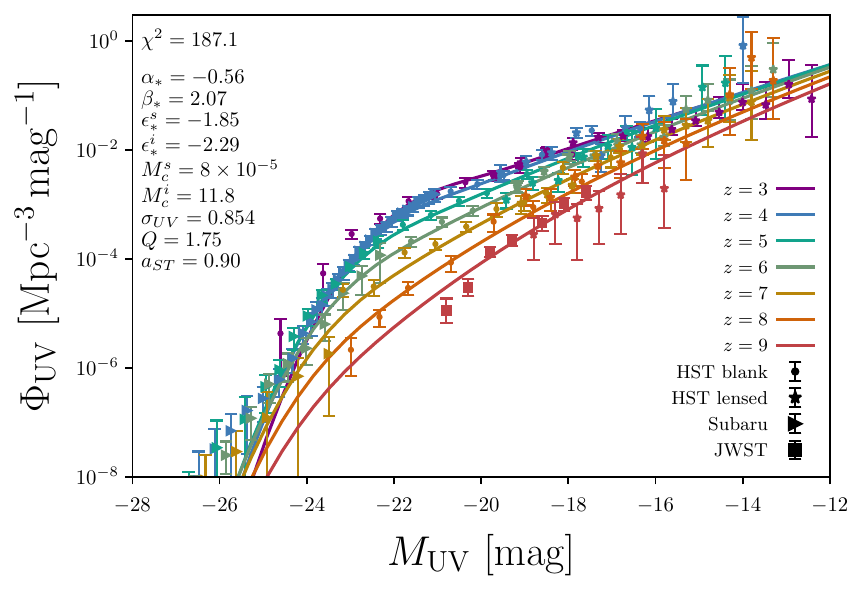}
    \caption{Combined UVLF dataset and CDM best-fit in our model. Corrections for dust and the Alcock-Paczy\'nski effect have been applied.}
    \label{figure:UVLF_data_and_bestfit}
\end{figure}

   $\lcdm$ provides a good fit to the UVLF data under consideration here, with a minimum $\chi^{2}$ of 187 and 173 degrees of freedom. The best-fit DAO parameters are degenerate with $\lcdm$, either with $f\ll 1$ or $\kpeak \gg 10\ h/\mathrm{Mpc}$. Figure \ref{figure:UVLF_data_and_bestfit} shows the combined dataset and best fit prediction within our model. 

\begin{figure}
    \centering
    \includegraphics[trim={0cm 0cm 0cm 0cm},clip,width=0.8\textwidth]{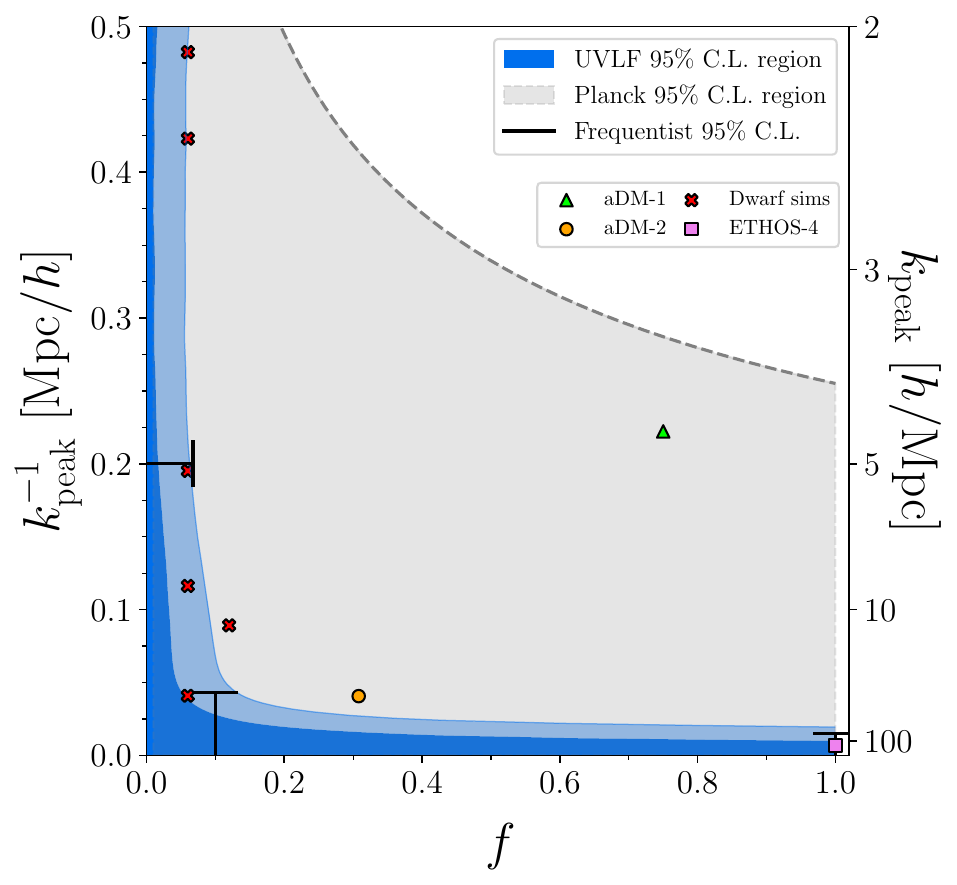}
    \caption{2D marginalized Bayesian constraints on the transfer function parameters $f$ and $1/\kpeak$, representing the fraction of DM experiencing DAOs and the dark sound horizon at decoupling, respectively. Constraints from CMB anisotropies are shown in gray, translated from constraints on atomic dark matter~\cite{Bansal:2022qbi}. Frequentist limits on $\kpeak$ obtained with profile likelihood scans for $f=0.1,\ 1$ and $\kpeak = 5\ h/\mathrm{Mpc}$ are shown in black. The projections of several specific model choices onto this parameter space are shown. aDM-1 corresponds to $\fd=0.75$, $\xid=0.3$, $\mpd=m_{p}$, $\med=4m_{e}$, $\alphad = 2\alpha$, and is consistent with CMB data but excluded by UVLF measurements. aDM-2 is the parameter point used to generate the example transfer function of Figure \ref{figure:adm_tk}, and is just ruled out by UVLF measurements. ETHOS-4 is an effective model within the ETHOS framework which has been studied in previous work, including in the context of the luminosity function~\cite{Vogelsberger:2015gpr,Lovell:2017eec}. It is allowed by UVLF measurements, consistent with the findings in \cite{Lovell:2017eec}. Finally, the aDM parameter choices for which hydrodynamical simulations of dwarf galaxies have been performed are shown~\cite{Roy:2024bcu}. Different choices of the dark photon temperature can lead to $\kpeak$ values spanning a wide range without significantly affecting the galactic-scale cooling physics.
    \label{figure:dao_constraints}}
    \end{figure}

    We show in Figure \ref{figure:dao_constraints} the 68\% and 95\% confidence level contours for the non-trivially constrained DAO parameters $f$ and $\kpeak$. For $\kpeak \gtrsim50\ h/\mathrm{Mpc}$, $f$ up to 1 is permitted, with the bound on $\kpeak$ gradually strengthening for larger $f$. Smaller values of $\kpeak$ are ruled out unless $f\lesssim 0.07$. Models with higher $\kpeak$ must decouple earlier, usually as a consequence of weaker interactions in the dark sector. For atomic dark matter, this can be achieved through a higher dark hydrogen binding energy or colder dark photon temperature. For dark matter scattering with baryons, lower cross-sections and higher dark matter masses achieve a similar effect. 
    
    At high $f$, our constraints improve significantly on re-cast bounds derived for atomic dark matter from CMB data, with the $\kpeak$ wave number restricted to be approximately an order of magnitude larger by UVLF measurements. However, for $\kpeak \lesssim 1\ h/\mathrm{Mpc}$, the lower experimental and modeling uncertainties of the CMB give it the advantage in constraining the abundance of dark matter undergoing DAO, permitting only $f\lesssim 5\%$. 

\begin{figure}[h]
    \centering
    \includegraphics[trim={0cm 0cm 0cm 0cm},clip,width=0.95\textwidth]{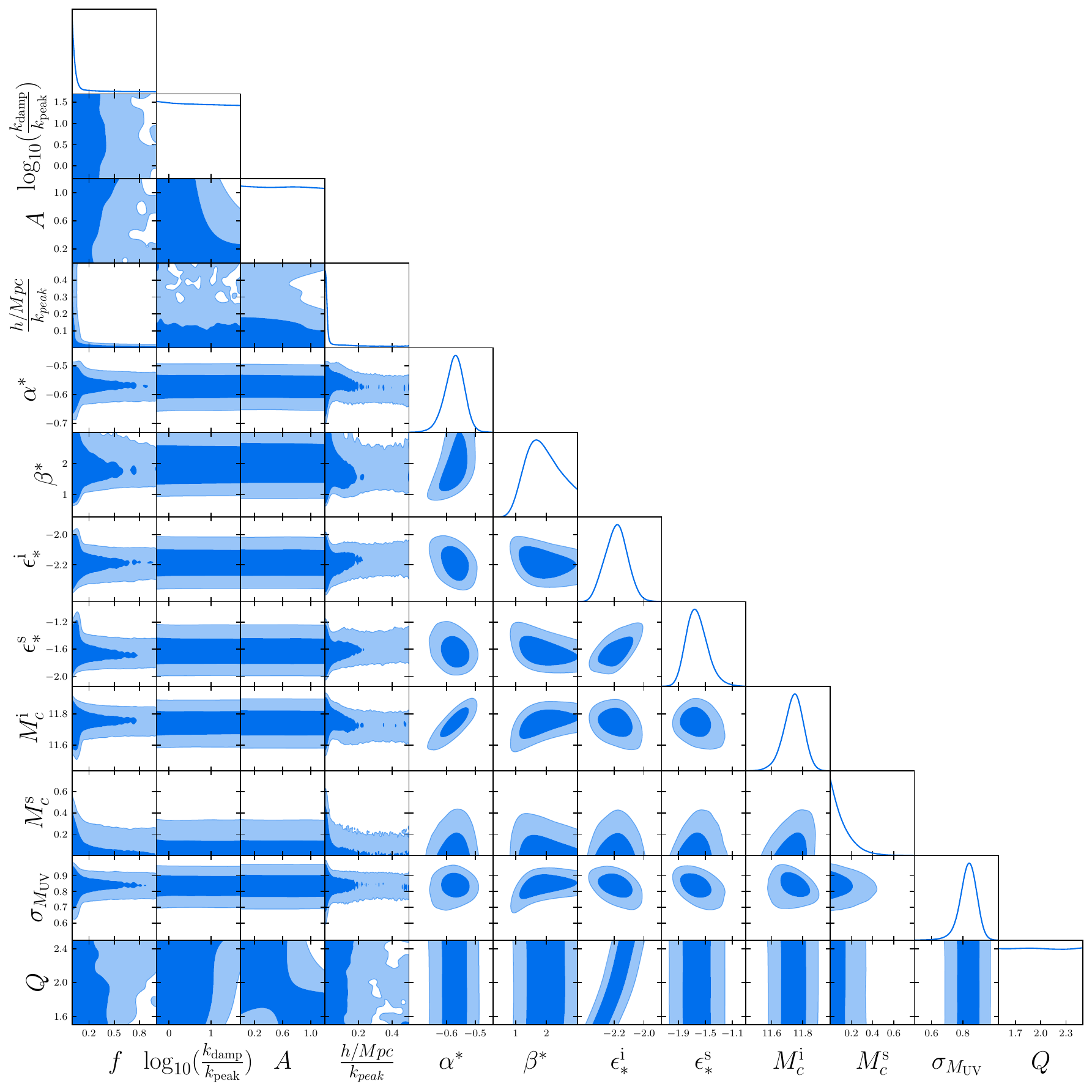}
    \caption{2D marginalized constraints on all model parameters. It is clear that neither $A$ nor $\kdamp/\kpeak$ are constrained when marginalizing over all other parameters.  
    \label{figure:adm_fullconstraints}}
    \end{figure}

    These UVLF observations have no constraining power over the DAO amplitude $A$ or the ratio of the damping scale to DAO scale, $\kdamp/\kpeak$. Both parameters have flat posteriors when marginalized over all other model parameters, and weak 2D correlations with other DAO parameters. This is likely because, while the dependence of the UVLF on $A$ can be large, it manifests at fainter magnitudes than the deviation from CDM for fixed $f$ and $\kpeak$, so that the lower bound on $\kpeak$ is driven by the initial suppression of the matter power spectrum. Our constraints can therefore be compared to those set on mixed warm/cold dark matter~\cite{Boyarsky:2008xj,Tan:2024cek}. We show the 2D marginalized posteriors for all DAO and astrophysical nuisance parameters in Figure \ref{figure:adm_fullconstraints}.
    
\begin{figure}
    \centering
    \includegraphics[trim={0cm 0cm 0cm 0cm},clip,width=0.8\textwidth]{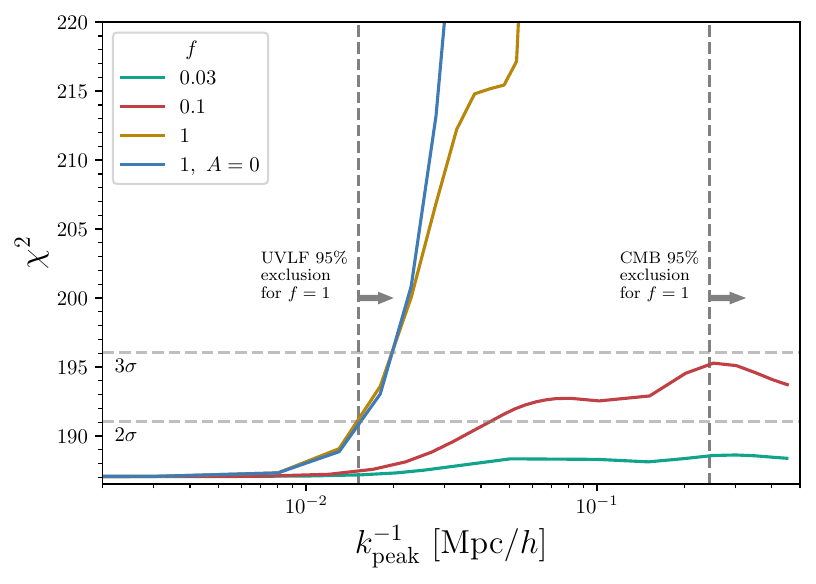}
    \caption{Profile likelihoods of $\kpeak^{-1}$ for $f=0.03,\ 0.1,\  1$, with $A=1$ and $\kdamp=10\ \kpeak$ fixed. A profile likelihood for $f=1$, $A=0$ is also shown. Bounds from the CMB and UVLF on $\kpeak$ for $f=1$ at 95\% confidence level are shown.
    \label{figure:prof_llh}}
    \end{figure}

    As an alternative to our Bayesian results, we present frequentist profile likelihoods, holding fixed either $f$ or $\kpeak$. For a few choices of $f$, we find the maximum likelihood as a function of $\kpeak$, marginalized over all nuisance parameters. We fix $A=1$ and $\kdamp=10\kpeak$ because they have flat posteriors in the Bayesian scan and exhibit little correlation with other parameters. This restricts the scan to the regime of strong DAOs. We perform the profile likelihood scan with the publicly available code Procoli~\cite{Karwal:2024qpt,Lewis:2019xzd,Audren:2012wb,Brinckmann:2018cvx}. Figure \ref{figure:prof_llh} displays the profile likelihoods for $f=0.03,\ 0.1,\ 1$. Consistent with the results of the Bayesian scan, the $\lcdm$ limit yields the best fit to the data. Using $\Delta \chi^{2}=4$ as a threshold to set a 95\% C.L. bound on $\kpeak$, we find that for $f=1$, $\kpeak > 66\ h/\mathrm{Mpc}$ and for $f=0.1$, $\kpeak > 23\ h/\mathrm{Mpc}$. We find no constraint on $\kpeak$ for $f=0.03$. The $2\sigma$ bound on $\kpeak$ for $f=1$ does not depend on $A$, although at lower $\kpeak$ the minimum $\chi^{2}$ is much smaller for $A=1$ than $A=0$. We also perform a profile likelihood scan over $f$ with $\kpeak$ fixed to $5\ h/\mathrm{Mpc}$, and derive a 95\% confidence level limit of $f < 0.07$. These frequentist limits are also displayed in Figure~\ref{figure:dao_constraints}. For an atomic dark sector, this means that only $\lesssim7\%$ of the dark matter can be atomic unless the dark photon bath is much colder and/or the dark hydrogen binding energy is much higher than in the SM. 
    
\begin{figure}
    \centering
    \includegraphics[trim={0cm 0cm 0cm 0cm},clip,width=0.8\textwidth]{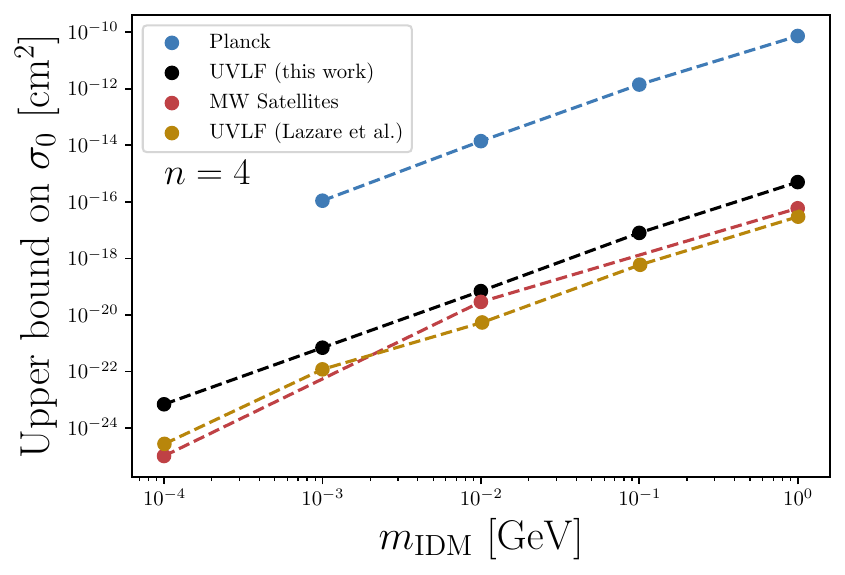}
    \caption{Re-cast 95\% C.L. constraints on the dark matter-baryon elastic scattering cross-section for a model with $n=4$ velocity dependence of the cross-section $\sigma=\sigma_{0}v^{n}$. We also show constraints from the CMB~\cite{Li:2022mdj}, estimated constraints from Milky Way satellite observations~\cite{Nadler:2025fcv}, and constraints using UVLF data computed for this specific model in \cite{Lazare:2025gha}.} 
    \label{figure:dmb_constraints}
    \end{figure}

    Because of the generality of our DAO transfer function model and the simplicity of the bounds in terms of $f$ and $\kpeak$, we can easily re-cast our bounds for specific models where $\kpeak$ can be computed. As a demonstrative example, 
    we re-cast the bound on $\kpeak$ for $f=1$ into a bound on $\sigma_{0}$ in the IDM model given a velocity dependence $n$ and dark matter mass $m_{\mathrm{IDM}}$. Using \texttt{CLASS}, we can compute the matter power spectrum in this model, fit our $T(k)$ model to the result, and find the $\sigma_{0}$ that gives $\kpeak\approx66\ h/\mathrm{Mpc}$. We perform this procedure for $n=4$ and show the results in Figure~\ref{figure:dmb_constraints} along with constraints from the CMB and Milky Way satellite abundances~\cite{Li:2022mdj,Nadler:2025fcv}. The MW satellite constraints are estimated by mapping IDM transfer functions to similar WDM transfer functions and applying the WDM bound. The UVLF constraint on the DM-baryon scattering cross-section is over five orders of magnitude stronger than the CMB constraint, and competitive with the MW satellite bound. A recent dedicated analysis of UVLF constraints on this model found somewhat stronger bounds, with limits improving upon CMB and satellite constraints~\cite{Lazare:2025gha}. The disparity between our results likely originates from our choice to use a different, more conservative measurement of the UVLF with HST lensed fields~\cite{2022ApJ...940...55B}.

    These results are complementary to existing work using Lyman-$\alpha$ forest measurements to constrain non-standard dark matter interactions in many different models which overlap with the DAO signature to varying extents. These include studies of warm dark matter, mixed warm/cold dark matter, strongly self-interacting dark matter, feebly interacting dark matter, ETHOS models, and dark matter-baryon interactions~\cite{Boyarsky:2008xj,Baur:2017stq,Irsic:2017ixq,Murgia:2017lwo,Krall:2017xcw,Murgia:2018now,Garny:2018byk,Bose:2018juc,Palanque-Delabrouille:2019iyz,Garzilli:2019qki,Archidiacono:2019wdp,Hooper:2022byl,Chatterjee:2023mlh,Verwohlt:2024efh,Das:2025bnr}. As a concrete example, for 100\% ETHOS dark matter with $n=4$ and weak DAOs, Lyman-$\alpha$ forest data from HIRES/MIKE were used to set a bound corresponding to $\kpeak > 136\ h/\mathrm{Mpc}$ in our model~\cite{Archidiacono:2019wdp}. This work mapped the weak DAO transfer function onto a WDM transfer function with the $\alpha\beta\gamma$ model to derive their limits~\cite{Murgia:2017lwo}. Because strong DAOs have been shown to impact the Lyman-$\alpha$ flux spectrum distinctly from WDM, it is non-trivial to translate this to a bound on strong DAOs with the same $\kpeak$~\cite{Bose:2018juc}. There is ongoing work to obtain robust Lyman-$\alpha$ forest constraints on strong DAOs.    
    
    \section{Conclusions}
    \label{sec:Conclusion}
    
    The space of dark matter models beyond the cold and collisionless paradigm is vast and immensely varied in possible phenomenology. To explore this space and identify which models are consistent with our Universe requires bringing to bear cosmological and astrophysical probes across many different scales and redshifts. It is also necessary to simultaneously consider multiple models in terms of signatures they have in common. In this work we have obtained novel constraints on dark matter models that exhibit dark acoustic oscillations at small scales by comparing with observations of the UV luminosity function at high redshifts by the Hubble Space Telescope, James Webb Space Telescope, Subaru Telescope, and Canada-France-Hawaii Telescope. 
    
    We introduced a simple analytical model for the DAO transfer function, with parameters that can easily be connected to physical quantities for a specific dark matter model, including the fraction of dark matter undergoing DAOs as well as the dark sound horizon. 
    
    Our model of the UV luminosity function uses the extended Press-Schechter formalism to predict the halo mass function from the linear matter power spectrum, calibrated against N-body simulations, and a highly flexible phenomenological model of the halo-galaxy connection whose parameters we marginalize over.

    From a Bayesian MCMC analysis allowing all parameters of the model to vary, we find that the transfer function parameter $f$ is constrained to be less than $\approx 0.07$ unless the wave number $\kpeak$ at which the DAOs appear is larger than $\approx 50\ h/\mathrm{Mpc}$. In the limit of large $\kpeak$, values of $f$ up to 1 are allowed. Allowing all parameters of the model to vary, no constraint is placed on the amplitude or damping of the DAOs. 
    
    From a frequentist profile likelihood analysis, we found that if $\kpeak> 66\ h/\mathrm{Mpc}$, then strong DAOs with $f=1$ are consistent with UVLF observations at the 95\% confidence level, and confirmed that for $f\lesssim 0.07$, all $\kpeak$ are allowed. 
    
    Because $f$ and $\kpeak$ are closely correlated with the fraction of dark matter that is not cold and collisionless and the dark sound horizon scale $\kdao$, respectively, we can apply our results to these physical parameters and set bounds on specific particle physics models. In particular, the bound on $f\lesssim 0.07$ can be straightforwardly interpreted as an upper limit on the fraction of dark matter that is atomic for $\kdao \lesssim 5\ h/\mathrm{Mpc}$. 
    
    We can also compare to previous bounds which were obtained for atomic dark matter using the CMB power spectrum. The most recent analysis of Planck data found that $\kdao > 3
    \ h/\mathrm{Mpc}$ \cite{Bansal:2022qbi} for $f = 1$. This is representative of the strength of the bound on strong DAOs in general from the CMB. Our results improve on the large-scale cosmological constraints by over an order of magnitude for fractional abundances over $\approx 10\%$. By strengthening the lower bound on $\kdao$, the epoch at which dark matter had to decouple from other species is pushed to earlier times. For atomic dark matter, this means reducing the dark photon temperature or increasing the dark hydrogen binding energy. An analysis of Lyman-$\alpha$ forest data for weak DAOs in the ETHOS formalism found that $\kpeak > 135\ h/\mathrm{Mpc}$, but did not extend their constraints to the strong DAO regime or fractional abundances of DM less than one~\cite{Bohr:2020yoe}. The strength of this result motivates further work to compute robust bounds on the strong DAO scenario with Lyman-$\alpha$ forest data.  

    Other probes on intermediate scales can furnish complementary constraints to these, as well as galactic-scale and astrophysical probes that are more directly sensitive to the dark matter interactions beyond the matter power spectrum. Studies of the effects of dissipation in the dark sector on galactic morphology with hydrodynamical simulations have shown that even small fractional abundances of dissipative dark matter may be strongly ruled out if cooling is efficient~\cite{Roy:2023zar,Roy:2024bcu}. For a model such as atomic dark matter, this constrains the viable parameter space in a way that can be completely orthogonal to bounds on $\kpeak$.  
    
    Even with our extremely flexible model of the halo-galaxy connection, we set new meaningful bounds on the DAO parameter space, demonstrating the utility of observables derived from higher redshift sources. As more data are collected with new and upcoming telescopes and our understanding of early galaxy formation evolves, we can expect that our ability to distinguish the more detailed features of the matter power spectrum will improve. Current JWST programs are delivering UVLFs at higher redshifts and fainter magnitudes than expected. These provide new, powerful constraints on the abundance of bright galaxies through wide surveys~\cite{2025arXiv250804791F}, as well as detections of the faintest objects yet with lensed surveys~\cite{Kokorev25_Glimpse}. A complete picture of galaxy formation at $z\gtrsim 10$ has not yet emerged, and only a handful of galaxies have been spectroscopically confirmed at $z\sim 14$~\cite{Naidu:2025xfo,Carniani24_GSz14}, but the high-$z$ regime promises to provide new insights on the clustering of dark matter.

    Upcoming telescopes like the Vera Rubin Observatory, Nancy Grace Roman Space Telescope, and Square Kilometer Array will collect a wealth of data that will undoubtedly transform our understanding of the history of our Universe and the nature of dark matter. In the absence of a direct or indirect detection of dark matter interactions with the Standard Model, it is imperative that we fully exploit the information we obtain to learn as much as we can from its gravitational interactions. 

    \section{Acknowledgments}

    We thank M. Lisanti for insightful comments and assistance in accessing computing resources. JB thanks D. Adams, A. Parikh, and L. Yuan for useful discussions. JB acknowledges support from NSF grants PHY-2210533 and PHY-2513893. The work of DC was supported in part by Discovery Grants from the Natural Sciences and Engineering Research Council of Canada (NSERC), the Canada Research Chair program, the Alfred P. Sloan Foundation, the Ontario Early Researcher Award, and the University of Toronto McLean Award.
    HL is supported by the U.S. Department of Energy under grant DE-SC0026297 and the Cecile K. Dalton Career Development Professorship, endowed by Boston University trustee Nathaniel Dalton and Amy Gottleib Dalton.
    JBM acknowledges support from NSF Grants AST-2307354 and AST-2408637, and the CosmicAI institute AST-2421782.
    The computations in this work were run at facilities supported by the Scientific Computing Core at the Flatiron Institute, a division of the Simons Foundation.

    \appendix\section{Alternative model of halo-galaxy connection}\label{sec:appendix}

    In the second model of the halo-galaxy connection that we use, instead of the star formation rate being related to the halo accretion rate, the average stellar mass in a galaxy is related to the halo mass by a double power law~\cite{Moster:2009fk}: 
    \begin{equation}
        \frac{M_{*}}{M_{\mathrm{h}}} = \frac{\epsilon_{*}}{\left( \frac{M_\mathrm{h}}{M_{c}}\right)^{\alpha_{*}} +\left( \frac{M_\mathrm{h}}{M_{c}}\right)^{\beta_{*}}} 
    \end{equation}
    $\epsilon_{*}$ and $M_{c}$ have the same redshift evolution as in the first model. 
    The stellar mass is assumed to be related to the star formation rate as~\cite{Park:2018ljd,Gillet:2019fjd,Munoz:2021psm}
    \begin{equation}
        M_{*} \propto \frac{\dot{M}_{*}}{H(z)}
    \end{equation}
    The constant of proportionality can be absorbed by $\epsilon_{*}$. The stellar accretion rate is then related to the UV luminosity by Equation \ref{eq:SFR-LUV}. 

\begin{figure}[h]
    \centering
    \includegraphics[trim={0cm 0cm 0cm 0cm},clip,width=0.8\textwidth]{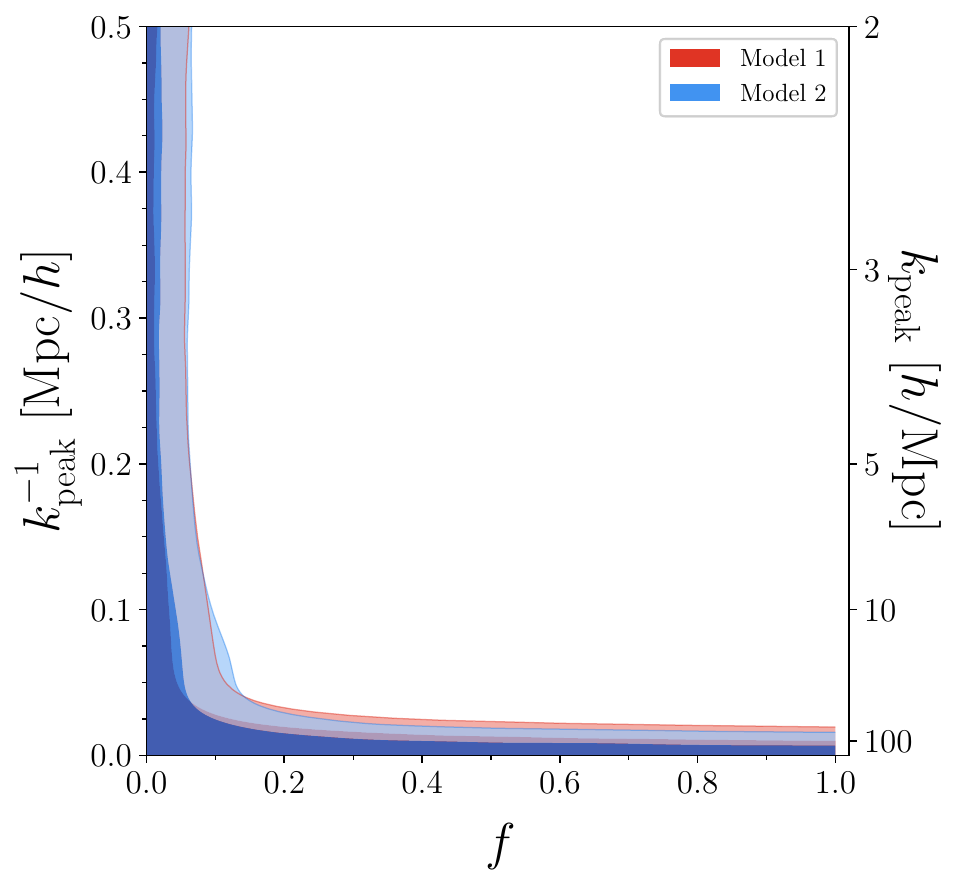}
    \caption{2D marginalized Bayesian constraints on the transfer function parameters $f$ and $\kpeak^{-1}$, representing the fraction of DM experiencing DAOs and the dark sound horizon at decoupling, respectively, for both models of the halo-galaxy connection.}
    \label{figure:model2_constraints}
    \end{figure}

    We show the constraints on $f$ and $\kpeak$ in Figure \ref{figure:model2_constraints}. Both models yield similar limits on the DAO parameters, with the limit on $\kpeak$ for high $f$ being slightly less strict for model 1. This motivates our choice to present results only for model 1 in the main body of the paper.

\bibliographystyle{jcap}
\bibliography{references.bib}
\end{document}